# Relaxation decoupling in metallic glassy state

P. Luo[1], P. Wen[1], H. Y. Bai[1], B. Ruta[2], and W. H. Wang[1]*

[1]Institute of Physics, Chinese Academy of Sciences, Beijing 100190, China.

[2]ESRF-The European Synchrotron, CS 40220, 38043 Grenoble Cedex 9, France.

*e-mail: whw@iphy.ac.cn

Upon cooling, glass-forming liquids experience a two-step relaxation associated to the cage rattling and the escape from the cage, and the following decoupling between the β- and the α-relaxations. The found decoupling behaviors have greatly changed the face of glassy physics and materials studies. Here we report a novel dynamic decoupling that the relaxation function changes gradually from a single-step to a two-step form as temperature declines through the stress relaxation of various metallic glasses in a broad time and temperature range below glass transition temperature ($T_g$). Such a two-step relaxation is unexpected in glassy state and reveals a decoupling of dynamic modes arising from two different mechanisms: a faster one exhibiting ballistic-like feature, and a slower one associated with a broader distribution of relaxation times typical of subdiffusive atomic motion. This first observation of two-step dynamics in metallic glassy state points to a far richer-than-expected scenario for glass relaxation.

The dynamic relaxation behavior in glassy systems is widely regarded as one of the most challenging unsolved problems in condensed matter physics and material science[1-5]. In the high temperature warm liquid, as temperature decreases a two-step decay of the intermediate scattering function appears[1,2]. This behavior can be attributed to two separated relaxation processes: the initial fast stage is associated to the trapping of the particles in the nearest neighbor cages before their escape, and the following slower process concerns the long range translational motion[1]. Upon cooling in the supercooled regime, there arises cooperative rearranging. In frequency domain, the liquid at sufficiently high temperature shows a single peak relaxation frequency, while as temperature is lowered to the moderately supercooled regime, the peak splits into fast (β) and slow (α) relaxations[1,3]. Further cool the supercooled liquid, if fast enough to avoid crystallization, a conceptually solid glass material can be obtained[3]. In current standard scenario, β-relaxation persists into the glassy state[6-8], while, the situation is less clear for the structural α-relaxation process.

Although the study of glass relaxation is made difficult owing to the non-equilibrium and non-ergodic nature, increasing efforts have been made recently which reveals that the dynamics of glass relaxation is different and far more complex as compared to their high-temperature precursors of supercooled liquid[9-14]. The existence of structural rearrangements far below $T_g$ has been recently reported in macroscopic studies of metallic glasses (MGs)[9,10] and a commercial silicate glass[11]. In the latter case, the macroscopic strain relaxation of the glass was attributed to the presence of mixed alkali ions in the network[11]. Subsequent measurements[12] of X-ray photon correlation spectroscopy (XPCS) in a single-alkali silicate glass broaden and strengthen the results reported in ref. 11 by showing the direct experimental evidence of the spontaneous atomic motion in glasses, where the relaxation time is found to be surprisingly fast, ~100-1000 s, even hundreds of degrees below $T_g$, in contrasts with the general idea of an ultraslow dynamics in the deep glassy state. During the aging of polymers at relatively lower temperatures, the enthalpy recovery was found to experience a double-step evolution towards equilibrium, indicating the presence of two time scales for glass equilibration and revealing a complex dynamic scenario[13].

Among structural glasses, MGs have the simplest atomic structure, which is often compared with dense random packing of hard spheres[15]. The structural relaxation processes underlying aging in MGs were recently found to be intermittent on atomic level[14], contrary to the common assumption of a steady slowing down of the dynamics. These findings have drawn a broad interest to the issues of glass relaxation, but meanwhile leave open the question on how to further understand the complex dynamics.

In this work, we investigated the relaxation dynamics of various MGs by following the stress decay under a constant strain at different temperatures ranging from around $T_g$ down to the deep glassy state. Our studies cover a rarely explored time window spanning more than five decades. We found surprisingly that as the temperature decreases the relaxation separates gradually into two processes. The two distinct relaxation steps are adequately caught by a double Kohlrausch-Williams-Watts (KWW) function[1]. By comparing both the timescale and the activation energy we suggest that the observed initial fast step relaxation cannot be simply ascribed to β-relaxation, implying an unexplored relaxation mode related to a different mechanism in glass.

Figure 1 shows the stress relaxation of three typical MGs of $Zr_{44}Ti_{11}Cu_{10}Ni_{10}Be_{25}$ ($T_g$ = 621 K), $Zr_{50}Cu_{40}Al_{10}$ ($T_g$ = 693 K) and $La_{55}Ni_{20}Al_{25}$ ($T_g$ = 471 K) at different temperatures (See the Methods for more details about the stress relaxation experiment). Before stress relaxation experiments, the MGs were pre-annealed at 0.9 $T_g$ for 48 h (see Methods). A decoupling of relaxation into two steps is observed in three MGs investigated. At moderately high temperature around $T_g$, the stress $\sigma(t)$ decays in a single-step manner (curves to the left with cross symbols). However, this behavior changes when the temperature is decreased to 20~30 K below $T_g$ (right curves with open symbols). There we observe the formation of a shoulder for intermediate times (around 10 min). As temperature is lowered further, this shoulder becomes more and more conspicuous, and the two-step relaxation phenomenon is revealed, unexpectedly, in the deep glassy state. For the lowest temperatures, the stress decay still continues following the initial fast step, although very slow.

For the single-step decay, the normal KWW function:

$$\sigma(t)/\sigma(0) = \exp[-(\Gamma_0 t)^{\gamma_0}], \tag{1}$$

provides an excellent fit to the data (see left curves with cross symbols in Fig. 1), where $\Gamma_0$ is the relaxation rate, $\gamma_0$ is the exponent ant $t$ is time. For the rest data sets presented in Fig. 1 (open symbols), the two-step decay can be perfectly captured by a double KWW function:

$$\sigma(t)/\sigma(0) = A\exp[-(\Gamma_1 t)^{\gamma_1}] + (1-A)\exp[-(\Gamma_2 t)^{\gamma_2}] \ (\Gamma_1 > \Gamma_2), \tag{2}$$

where $\Gamma_1$ and $\Gamma_2$ represent the characteristic fast and slow relaxation rates, respectively; $\gamma_1$ and $\gamma_2$ are the corresponding exponents; prefactors $A$ and (1-$A$) manifest the relaxation strength.

The parameters obtained from fits to the stress relaxation functions in Fig. 1 are plotted in Fig. 2 as a function of $T_g/T$ for better comparisons. Figure 2a shows an obvious and universal decoupling between the fast and slow relaxation processes around $T_g/T \sim 1.03$ in the temperature dependent relaxation rates for all the three MGs. Both $\Gamma_1$ (half-filled symbols) and $\Gamma_2$ (open symbols) decrease with temperature, but their temperature dependence differs a lot. $\Gamma_2$ slows down for as large as ten decades, while $\Gamma_1$ varies within less than one decade over the probed temperature range. In addition, $\Gamma_2$ of three MGs collapse well onto a single master curve, implying the existence of a similar dynamics for the slower step. The Arrhenius fit to the data yields an activation energy $\triangle E_2 = (51.7 \pm 1.1) \ k_B T_g$, i.e., 2~3 eV for the investigated MGs. This collapse does not occur for $\Gamma_1$ when plotted against $T_g/T$ (see Fig. S1 for more clarity), while the data overlap well when plotted as a function of the inverse temperature (1000 K)/$T$ (the inset in Fig. 2a). The Arrhenius fit yields an activation energy $\triangle E_1$ around but less than 0.1 eV independent of $T_g$, or rather, the composition.

Figure 2b shows the temperature dependence of the KWW exponents. For the high temperature single-step relaxation, $\sigma(t)$ exhibits a stretched shape ($\gamma_0 < 1$), and $\gamma_0$ decreases with temperature. In the two-step regime the data can be described by two KWW decays with distinct exponents: as temperature decreases the exponent of the first decay, $\gamma_1$, (half-filled symbols) grows progressively from around 0.7 to above 1, finally saturating around 1.3 for both Zr-based MGs and around 1.5 for La-based MG. Conversely, the exponent of the second decay, $\gamma_2$, is less than 1, experiences a continuous decrease with temperature, and is independent of glass composition. This contrasting going-up versus going-down evolution of $\gamma_1$ and $\gamma_2$ with temperature decreasing further highlights the decoupling between the fast and slow relaxation mechanisms. Figure 2c shows that the strength *A* of the fast relaxation process decreases with temperature. At higher temperatures, the three MGs share a similar trend with *A* sharply decreasing with temperature; at low temperatures, *A* decreases faster for the La-based MG as compared to the other two.

Figure 3 plots the stress relaxation function of $Zr_{44}Ti_{11}Cu_{10}Ni_{10}Be_{25}$ under different strains at 519 K. The upper right inset in Fig. 3 shows the initial stress before relaxation as a function of the imposed strain. As the strain grows to around 1% the stress deviates from the linear dependence on strain, i.e., yielding starts. At small strains (below 0.7%) the normalized relaxation strength of the fast step grows with strain. In the plastic flowing regime (above 1%) the relaxation still follows a two-step behavior and the curves collapse. The fitted relaxation rates and exponents appear independent of the imposed strain within the experimental uncertainty (see the lower left inset in Fig. 3). Such a two-step behavior provides a new clue to understand the plastic deformation of MGs.

The relaxation spectra observed here, phenomenologically, bear a striking resemblance to what usually observed in the decay of the density fluctuations in Lennard-Jones glasses[2,16], molecular supercooled liquids[17] and colloidal suspensions[18-20]. In these works the fast process is independent of both temperature and time and is due to microscopic interactions between a particle and the cage created by its nearest neighbors. Its behavior reflects a diffusive particle motion and it is usually described by a single exponential decay (*exponent* =1). Differently from those works, here we probe the deep non-equilibrium glassy state, and most importantly, a new decoupling between fast and slow dynamics occurs with the relaxation evolving from the single-step to the two-step decay as temperature decreases. According to our current knowledge, two kinetic processes are responsible for glass relaxation i.e., the β-relaxation and the slow structural α-relaxation processes[6,7]. In the first sight, it seems plausible that the unexpected fast process observed here can be ascribed to β-relaxation. However, these two processes are incompatible with each other on both the timescale and the apparent activation energy. The β-relaxation is generally identified in the frequency domain by dielectric spectroscopy in non-MGs or by dynamical mechanical analysis in MGs[6]. Typically for MGs, the relaxation rate of β-relaxation is in the order of 1 s$^{-1}$ around $0.7T_g$ (refs 6, 7), while it is in the order of $10^{-3}$ s$^{-1}$ for the initial fast step (see Fig. 2a). In addition, the apparent activation energy of β-relaxation depends on $T_g$ and it would be approximately $26k_BT_g$ for MGs, i.e., 1.1~1.6 eV for the systems investigated in this work. As mentioned before, remarkably, the fast step reported here has almost negligible apparent activation energy of about 0.1 eV, which thus strongly contrasts with β-relaxation. Therefore, we conclude that the initial fast step observed here cannot be ascribed to the well-known β-relaxation. The initial fast processes herein also differ from that observed in enthalpy relaxation of glassy polymers[13] by showing huge discrepancy in both timescale and activation energy, suggesting a different microscopic mechanisms probably related to the different microscopic nature of the investigated glass systems. Notwithstanding, in both cases, the occurrence of an additional step during the glass relaxation implies the existence of a much complex dynamic behavior during structural relaxation than the simple steady evolution usually reported in macroscopic studies[15].

The peculiar behavior that $γ_1$ reaches above 1 for $T_g/T >1.1$ is in contrast with the following slower step and with previous studies[3-5] where the decay exponent is stretched, thus less than 1. Such anomalous compressed exponential relaxation has been observed in the microscopic dynamics of a variety of systems, including soft matters[21-28], MGs[14,29-31] and even magnetic structures[32], and it has been described in terms of a ballistic particle motion relaxing randomly distributed local stresses[21-23]. A similar explanation could fit also in the initial stage of the relaxation in our measurements. Before the slower subdiffusive motion characterized by the typical stretched exponential relaxation[4,22], ballistic-like particle motion driven by local stress dipoles proceeds, leading to a fast compressed exponential relaxation. On the other hand, the inset in Fig. 2a shows that $\Gamma_1$ depends only on temperature regardless of composition, although the temperature dependence is so weak with the activation energy less than 0.1 eV. This suggests that in the initial fast relaxation local internal stress dominates instead of thermal energy.

Although $\Gamma_1$ varies very little in the explored temperature range, the exponent $\gamma_1$ changes a lot especially in the intermediate temperature range (1.03<$T_g/T$<1.1) where the strength $A$ of the fast secondary process experiences a rapid decrease with temperature (see Fig. 2c). Similar phenomenon has also been seen in colloidal suspensions approaching jamming transition where the KWW exponent grows gradually from 0.86 and then saturates around 1.3 as the volume fraction increases[28]. In that sense, it seems plausible that the decrease of $\gamma_1$ is associated with the decrease in volume fraction as temperature increases. This is consistent with the picture of the local stress dipoles dominated ballistic-like particle motion preceding the long-range structural rearrangements. In MGs the volume increases by approximately 1.5% from room temperature to around $T_g$ due to thermal expansion[33], which is large enough to affect the dynamics for a molecular glass.

The relevant finding of the decoupling of relaxation dynamics accompanied by the occurrence of two-step decay in the mechanical response in glassy state extends our knowledge far beyond the orthodox understanding of glass relaxation, and a more complete scenario concerning the relaxation dynamics from high temperature liquid to glassy state can be sketched. As shown schematically in Fig.4, for high temperature liquids, as temperature decreases the relaxation dynamics of the system experiences the two-step behavior of the cage rattling and the escape from the cage, then the decoupling between the β- and the α-relaxations. When the system attains glassy state, we still see another decoupling of the relaxation modes. We suggest that the faster mode could be related to ballistic-like particle motion which arises from the local internal stresses annihilation, and the slower mode could be associated to a broader distribution of relaxation times typical of subdiffusive atomic motion. We distinguish the faster mode from the common β-relaxation by their dramatic difference in both timescales and apparent activation energy.

Distinct dynamical processes and intermittent structural relaxation have been recently observed also in the atomic motion of various glasses by means of XPCS[12,14,29-31]. Intriguingly, we note that the measured relaxation times occur on time scales comparable with that of the initial fast step in our data, and both processes share the analogous compressed exponential dynamics associated to internal stresses[12,14,29-31]. XPCS provides information on the density fluctuations on a microscopic scale of few Å (refs 29, 30). The second slower dynamical process observed in our data may therefore be absent as associated to larger length scales. Our findings suggest that further studies are deserved to explore the rich dynamics and in-depth mechanisms of glass relaxation.

**References**

1. Donth, E. *The Glass Transition* (Springer, Berlin, 2001).

2. Kob, W., Andersen, H. C. Testing mode-coupling theory for a supercooled binary Lennard-Jones mixture. II. Intermediate scattering function and dynamic susceptibility. *Phys. Rev. E* **52**, 4134-4153 (1995).

3. Debenedetti, P. G., Stillinger, F. H. Supercooled liquids and the glass transition. *Nature* **410**, 259-267 (2001).

4. Phillips, J. C. Stretched exponential relaxation in molecular and electronic glasses. *Rep. Prog. Phys.* **59**, 1133-1207 (1996).

5. Ediger, M. D. Spatially heterogeneous dynamics in supercooled liquids. *Annu. Rev. Phys. Chem.* **51**, 99-128 (2000).

6. Yu, H. B. *et al*. The β-relaxation in metallic glasses. *National Sci. Rev.* **1**, 429-461 (2014).

7. Luo, P. *et al*. Prominent β-relaxations in yttrium based metallic glasses. *Appl. Phys. Lett.* **106**, 031907 (2015).

8. Zhu, F. *et al*. Intrinsic correlation between b-relaxation and spatial heterogeneity in a metallic glass. *Nature Commun.* **7**, 11516 (2016).

9. Luo, P. *et al*. Probing the evolution of slow flow dynamics in metallic glasses. *Phys. Rev. B* **93**, 104204 (2016).

10. Lu, Z. *et al*. Flow unit perspective on room temperature homogeneous plastic deformation in metallic glasses. *Phys. Rev. Lett.* **113**, 045501 (2014).

11. Welch, R. C. *et al*. Dynamics of glass relaxation at room temperature. *Phys. Rev. Lett.* **110**, 265901 (2013).

12. Ruta, B. *et al*. Revealing the fast atomic motion of network glasses. *Nature Commun.* **5**, 3939 (2014).

13. Cangialosi, D. *et al*. Direct evidence of two equilibration mechanisms in glassy polymers. *Phys. Rev. Lett.* **111**, 095701 (2013).

14. Evenson, Z. *et al*. X-ray photon correlation spectroscopy reveals intermittent aging dynamics in a metallic glass. *Phys. Rev. Lett.* **115**, 175701 (2015).

15. Wang, W. H. The elastic properties, elastic models and elastic perspectives of

metallic glasses. *Prog. Mater. Sci.* **57**, 487-656 (2012).

16. Kob, W. *et al*. Aging effects in a Lennard-Jones glass. *Phys. Rev. Lett.* **78**, 4581 (1997).

17. Petzold, N. *et al*. Light scattering study on the glass former o-terphenyl. *J. Chem. Phys.* **133**, 124512 (2010).

18. Zheng, Z. Y. *et al*. Glass transitions in quasi-two-dimensional suspensions of colloidal ellipsoids. *Phys. Rev. Lett.* **107**, 065702 (2011).

19. Brambilla, G. *et al*. Probing the equilibrium dynamics of colloidal hard spheres above the Mode-Coupling glass transition. *Phys. Rev. Lett.* **102**, 085703 (2009).

20. de Melo Marques, F. A. *et al*. Structural and microscopic relaxations in a colloidal glass. *Soft Matter* **11**, 466-471 (2015).

21. Cipelletti, L. *et al*. Universal aging features in the restructuring of fractal colloidal gels. *Phys. Rev. Lett.* **84**, 2275-2278 (2000).

22. Bouchaud, J.-P. & Pitard, E. Anomalous dynamical light scattering in soft glassy gels. *Eur. Phys. J. E* **6**, 231-236 (2001).

23. Ferrero, E. E. *et al*. Relaxation in yield stress systems through elastically interacting activated events. *Phys. Rev. Lett.* **113**, 248301 (2014).

24. Guo, H. Y. *et al*. Nanoparticle motion within glassy polymer melts. *Phys. Rev. Lett.* **102**, 075702 (2009).

25. Aravinda Narayanan, R. *et al*. Dynamics and internal stress at the nanoscale related to unique thermomechanical behavior in polymer nanocomposites. *Phys. Rev. Lett.* **97**, 075505 (2006).

26. Orsi, D. *et al*. 2D dynamical arrest transition in a mixed nanoparticle-phospholipid layer studied in real and momentum spaces. *Sci. Rep.* **5**, 17930 (2015).

27. Bursac, G. *et al*. Cytoskeletal remodelling and slow dynamics in the living cell. *Nature Mater.* **4**, 557-561 (2005).

28. Ballesta, P. *et al*. Unexpected drop of dynamical heterogeneities in colloidal suspensions approaching the jamming transition. *Nature Phys.* **4**, 550-554 (2008).

29. Giordano, V. M. & Ruta, B. Unveiling the structural arrangements responsible for

the atomic dynamics in metallic glasses during physical aging. *Nature Commun.* **7**, 10344 (2016).

30. Ruta, B. *et al*. Atomic-scale relaxation dynamics and aging in a metallic glass probed by X-ray photon correlation spectroscopy. *Phys. Rev. Lett.* **109**, 165701 (2012).

31. Ruta, B. *et al*. Compressed correlation functions and fast aging dynamics in metallic glasses. *J. Chem. Phys.* **138**, 054508 (2013).

32. Shpyrko, O. G. *et al*. Direct measurement of antiferromagnetic domain fluctuations. *Nature* **447**, 68-71 (2007).

33. Yavari, A. R. *et al*. Excess free volume in metallic glasses measured by X-ray diffraction. *Acta Mater.* **53**, 1611-1619 (2005).

**Methods**

**MGs preparation.** Master alloys were prepared by arc-melting elements with purities above 99.99%. MG ribbons with 25 μm thick were produced by melt-spinning method under argon atmosphere. The glassy nature was ascertained by x-ray diffraction (XRD) in a Bruker D8 AA25 diffractometer with $Cu\ K_\alpha$ radiation. Differential scanning calorimetry was performed to determine $T_g$, in a Perkin-Elmer DSC 8000 at a heating rate of 20 K/min.

**Pre-annealing.** MGs are usually produced by rapid quench where the cooling rate can be as large as $10^6$ K/s, then internal stresses are inevitably built in as well as large scale structural heterogeneity[34,35]. Such heterogeneity may probably act as an extrinsic effect that masks possible intrinsic dynamics[4,9]. To rule out the possible influence of the large as-cast heterogeneity, we anneal the MG, before the stress relaxation experiments, in a muffle furnace with temperature stability of ±2 K. The samples were sealed in quartz tubes filled with high-purity argon gas to prevent oxidation.

**Stress relaxation experiment.** This was conducted on a dynamic mechanical analyzer (TA DMA Q800). As the temperature was stable tensile strain was applied on the ribbon and the stress evolution was simultaneously detected. After the stress relaxation the samples remain glassy as examined by XRD.

34. Wagner, H. *et al*. Local elastic properties of a metallic glass. *Nature Mater.* **10**, 439-442 (2011).
35. Liu, Y. H. *et al*. Characterization of nanoscale mechanical heterogeneity in a metallic glass by dynamic force microscopy. *Phys. Rev. Lett.* **106**, 125504 (2011).

**Acknowledgements**

We are indebted to W. Kob, M. Ballauff, M. Fuchs, T. Voigtmann, Y. Z. Li, Y. T. Sun, H. L. Peng, M. Z. Li and Y. L. Han for very helpful discussions. This research was supported by the NSF of China (51271195) and MOST 973 Program (No. 2015CB856800).

**Author Contributions** P.L. carried out the experiments. W.H.W. led the project and supervised the research. P.L., P.W. and W.H.W. analyzed the data and wrote the paper. All authors contributed to comment on the manuscript writing and the result discussions.

**Additional Information**

Correspondence and requests for materials should be addressed to W.H.W.

**Competing financial interests**

The authors declare no competing financial interests.

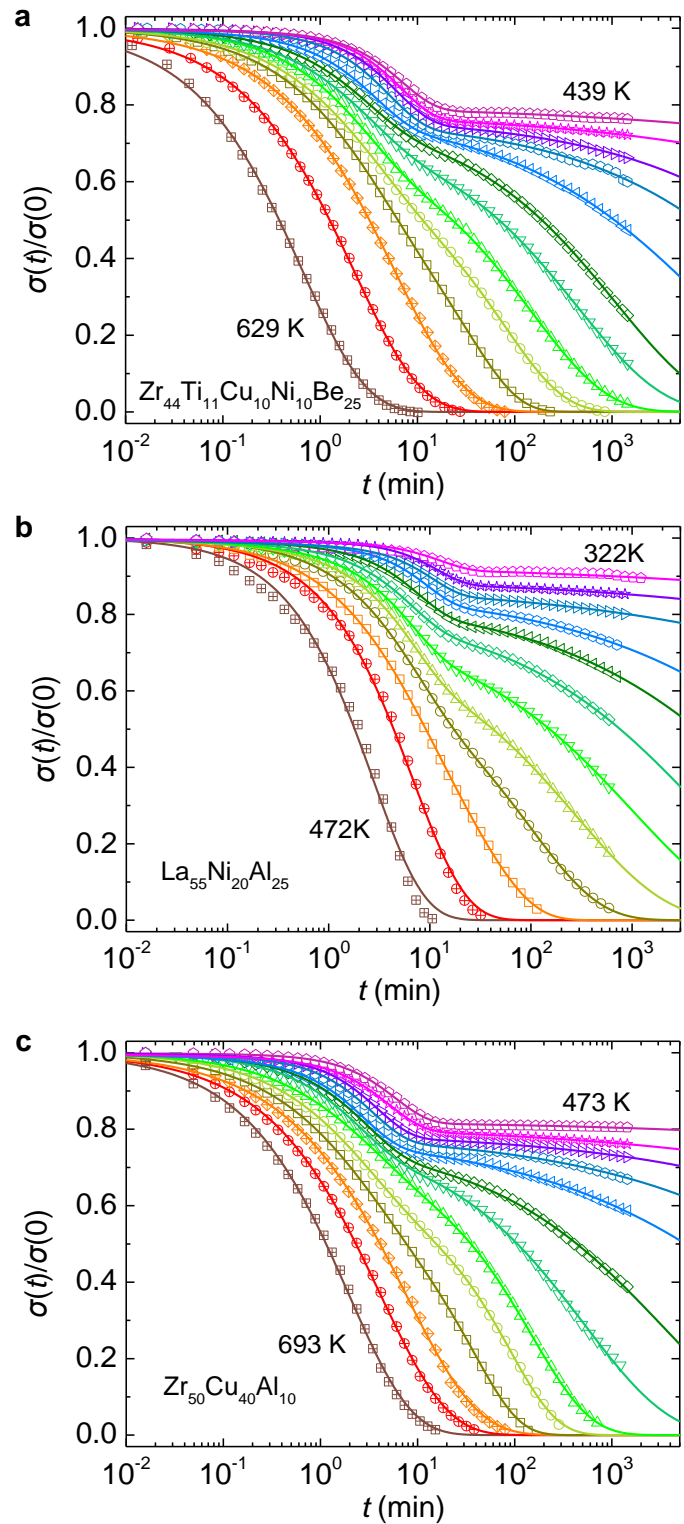

**Figure 1 | Temperature dependence of stress relaxation of three MGs. a**, $Zr_{44}Ti_{11}Cu_{10}Ni_{10}Be_{25}$ MG, from bottom to top and left to right, the temperatures

investigated are $T$ = 629, 619, 609, 599, 589, 579, 569, 559, 539, 519, 499, 469 and 439 K. **b**, $La_{55}Ni_{20}Al_{25}$ MG, and $T$ = 472, 462, 452, 442, 432, 422, 412, 402, 382, 362, 342 and 322 K. **c**, $Zr_{50}Cu_{40}Al_{10}$ MG, and $T$ = 693, 683, 673, 663, 653, 643, 623, 603, 583, 563, 533, 503 and 473 K. The strain is 0.3%. For the sake of clarity, the stress $\sigma(t)$ has been normalized by its initial value $\sigma(0)$ at $t$ = 0. Solid lines are theoretical fits to the data. The obtained fitting parameters are plotted in Fig.2. See the text for further details.

**Figure 2**

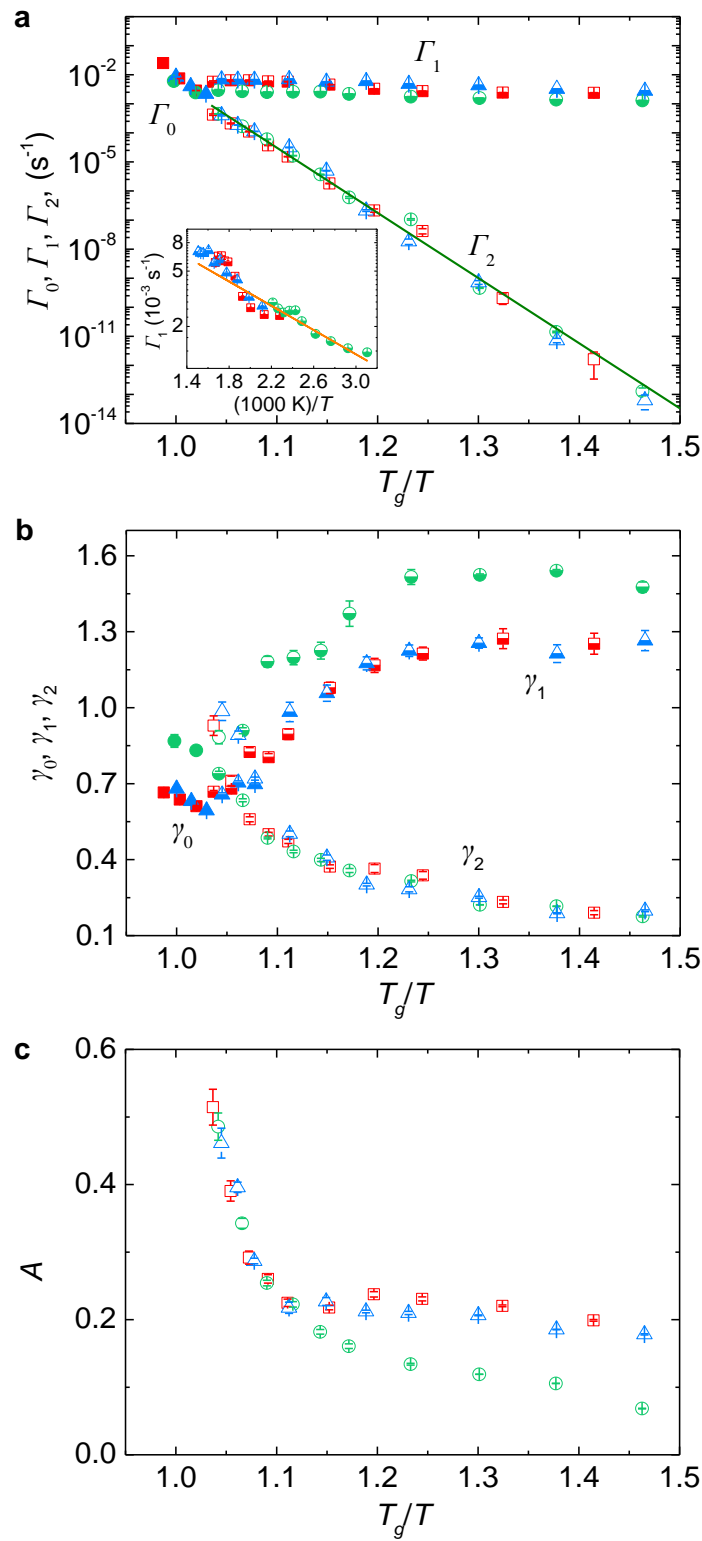

**Figure 2 | Temperature dependent relaxation dynamics. a-c**, Temperature dependence of the relaxation rate (**a**), exponent (**b**) and prefactor (**c**), obtained from

the fitting of equation (1) to the high temperature data (cross symbols) and equation (2) to the lower temperature data (open symbols) in Fig. 1. Symbols: filled symbols represent $\Gamma_0$ and $\gamma_0$, half-filled symbols represent $\Gamma_1$ and $\gamma_1$, and open symbols represent $\Gamma_2$, $\gamma_2$ and $A$; red squares refer to $Zr_{44}Ti_{11}Cu_{10}Ni_{10}Be_{25}$ MG, green circles are $La_{55}Ni_{20}Al_{25}$ MG, and blue triangles are $Zr_{50}Cu_{40}Al_{10}$ MG. The inset in panel **a** shows that $\Gamma_1$ of the three samples collapse onto one master curve when plotted against (1000 K)/T. Solid lines in panel **a** and the inset are the best linear fit to the data. Error bars represent the range of parameters obtained from fitting the stress relaxation data in Fig. 1. Where not shown, error bars are within the symbol size.

**Figure 3**

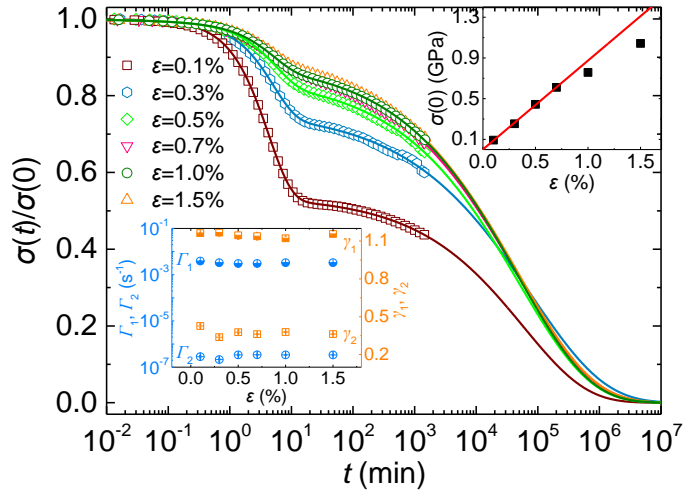

**Figure 3 | Strain dependence of the relaxation dynamics.** Stress relaxation of $Zr_{44}Ti_{11}Cu_{10}Ni_{10}Be_{25}$ MG at 519 K with different strains $\varepsilon$, solid lines are the fitting of equation (2) to the data. The lower left inset shows the fitted characteristic relaxation rate and exponent as a function of strain. The upper right inset plots the initial stress $\sigma(0)$ as a function of strain, yielding starts around $\varepsilon=1\%$.

**Figure 4**

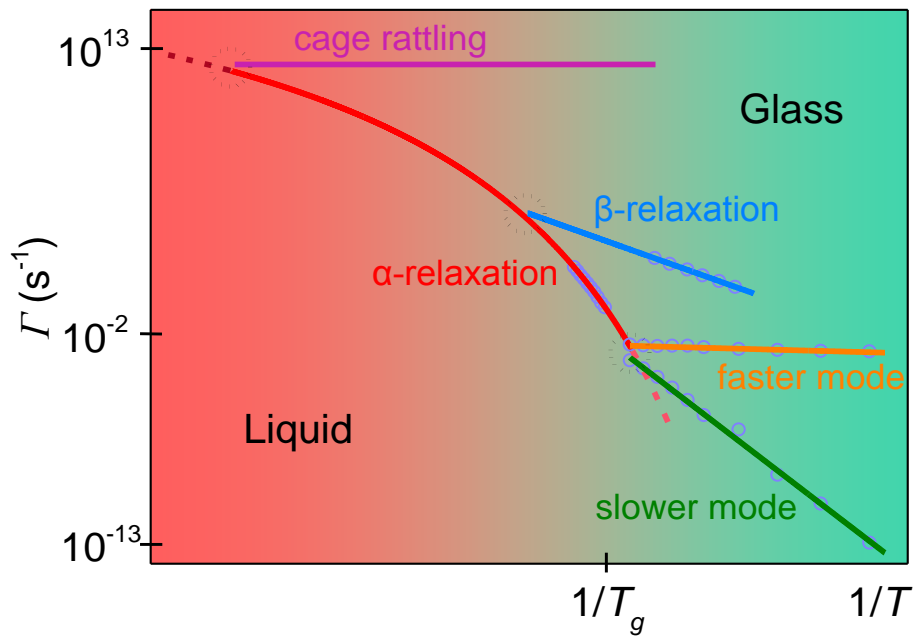

**Figure 4 | A schematic Arrhenius diagram concerning the dynamical behaviors of MG and its high temperature precursors.** Data for the α- and the β-relaxations are from dynamical mechanical measurements on $La_{55}Ni_{20}Al_{25}$ MG (Fig. S3). The red line is fit of the Vogel-Fulcher-Tammann (VFT) equation[3] to the α-relaxation data and the blue line is fit of the Arrhenius equation to the β-relaxation data.

## Supplementary Information

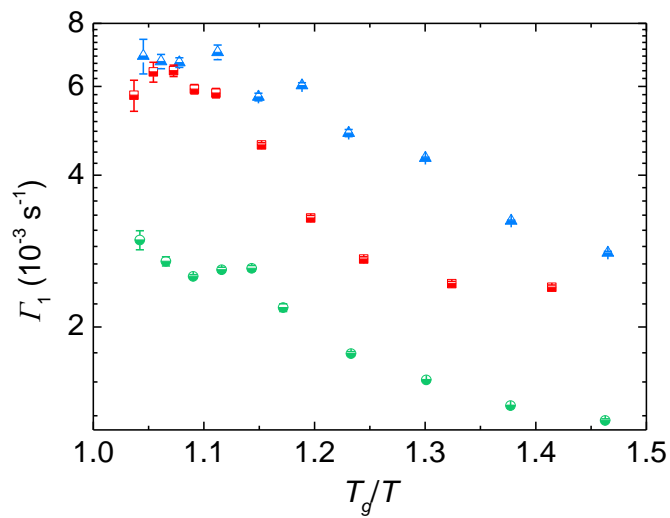

**Figure S1 | The initial fast relaxation rate $\Gamma_1$ plotted against $T_g/T$.** Red squares refer to $Zr_{44}Ti_{11}Cu_{10}Ni_{10}Be_{25}$ MG, green circles refer to $La_{55}Ni_{20}Al_{25}$ MG, and blue triangles refer to $Zr_{50}Cu_{40}Al_{10}$ MG.

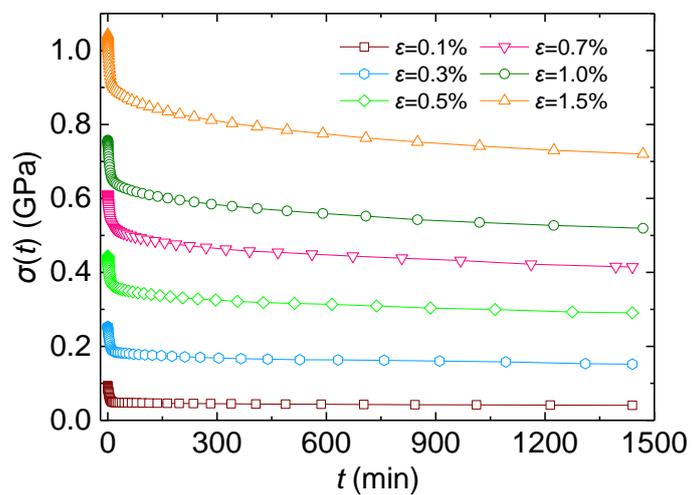

**Figure S2 | Strain dependence of the stress decay functions in $Zr_{44}Ti_{11}Cu_{10}Ni_{10}Be_{25}$ MG at 519 K.** As shown in the inset of Fig. 3 in the main text, the initial stress $\sigma(0)$ reveals that yielding starts around $\varepsilon=1\%$.

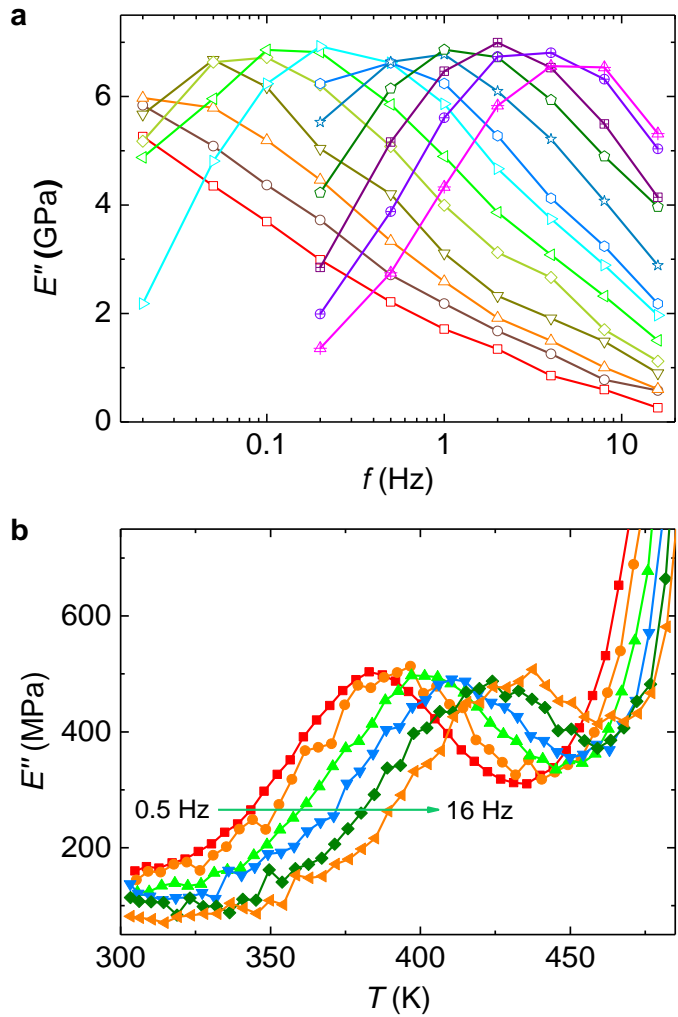

**Figure S3 | The α- and β-relaxation spectra for La$_{55}$Ni$_{20}$Al$_{25}$ MG.** (a) Frequency dependent loss modulus above $T_g$ reveals the α-relaxation, from left to right the temperature increases from 463 K to 499 K with a 3 K interval. (b) Temperature dependent loss modulus at different frequencies, from left to right: 0.5, 1, 2, 4, 8, and 16 Hz, at a heating rate of 3 K/min. The sample was pre-annealed at $0.9T_g$ for 48 h. The β-relaxation appears as a peak in the loss modulus. The relaxation rate for the α- and β-relaxations is $2\pi f_{peak}$, where $f_{peak}$ is the peak frequency. The measurements are performed on TA DMA Q800.